\documentclass[preprint]{aastex631}

\usepackage{subeqnarray}
\usepackage{amsmath}
\usepackage{hyperref}
\shorttitle{the general relativistic spin precession}
\shortauthors{Yuan et al.}
\graphicspath{{./}{figures/}}

\begin{document}

\title{The Influence of General Relativity on the Spins of Celestial Bodies in Inclined Orbits}

\author{Huan-rong Yuan}
\affiliation{School of Mathematics, Physics and Statistics Shanghai
University of Engineering Science, \\
Shanghai 201620, China\\}

\author[0000-0003-0506-054X]{Ying Wang}
\affiliation{School of Mathematics, Physics and Statistics Shanghai
University of Engineering Science, \\
Shanghai 201620, China\\}
\email{wangying424524@163.com}

\author[0000-0003-2620-6835]{Xin Wu}
\affiliation{School of Mathematics, Physics and Statistics Shanghai
University of Engineering Science, \\
Shanghai 201620, China\\}

\author[0000-0002-6472-5348]{Ji-wei Xie}
\affiliation{School of Astronomy and Space Science and Key Laboratory of Modern Astronomy and Astrophysics in Ministry of Education,\\
 Nanjing University, Nanjing 210093, China\\}

\author[0000-0001-5162-1753]{Hui-gen Liu}
\affiliation{School of Astronomy and Space Science and Key Laboratory of Modern Astronomy and Astrophysics in Ministry of Education,\\
 Nanjing University, Nanjing 210093, China\\}

\author[0000-0003-1680-2940]{Ji-lin Zhou}
\affiliation{School of Astronomy and Space Science and Key Laboratory of Modern Astronomy and Astrophysics in Ministry of Education,\\
 Nanjing University, Nanjing 210093, China\\}

\author[0000-0003-2620-6835]{Wei Sun}
\affiliation{School of Mathematics, Physics and Statistics Shanghai
University of Engineering Science, \\
Shanghai 201620, China\\}



\begin{abstract}
Through the Rossiter-McLaughlin effect, several hot Jupiters have been found to exhibit spin-orbit misalignment, and even retrograde orbits. The high obliquity observed in these planets can be attributed to two primary formation mechanisms, as summarized in the existing literature. First, the host star's spin becomes misaligned with the planetary disk during the late stages of star formation, primarily due to chaotic accretion and magnetic interactions between the star and the planetary disk. Second, the orbital inclination of an individual planet can be excited by dynamical processes such as planet-planet scattering, the Lidov-Kozai cycle, and secular chaos within the framework of Newtonian mechanics. This study introduces a third mechanism, where, within the framework of general relativity, the post-Newtonian spin-orbit coupling term induces precession of the host star's spin around the orbital angular momentum. The orbital inclination, relative to a reference plane, can expand the range of deviation in the spatial orientation of the bodies' spins from the plane's normal. The varying amplitude and period of spin precession for both the star and the planet are derived theoretically, and the results, which can be applied without restriction, agree well with numerical simulations.

\end{abstract}



\section{Introduction} \label{sec:intro}

After centuries of research and observation by astronomers, the theory of planetary formation within our solar system is generally well-established. However, a significant challenge arose in 1995 when Mayor and Queloz announced the discovery of the first exoplanet orbiting a Sun-like star, 51 Pegasi b \citep{Mayor1995}. This planet, with a mass of 0.46 $M_{\rm J}$ (where $M_{\rm J}$ is the mass of Jupiter), has an exceptionally short orbital period of 4.231 days, making it the first "hot Jupiter" observed. Planet formation theories include the core accretion model \citep{Zahn1977} and the gravitational instability model \citep{Toomre1964}. According to the core accretion model, solid cores of giant planets typically form several astronomical units (au) away from their host star, where sufficient raw material is available. This makes it difficult for hot Jupiters to acquire enough material to form in situ. As a result, it is generally believed that gas giants form at a greater distance from their host star and later migrate inward, either through gas disk migration \citep{Goldreich1980, Lin1986, Lin1996, Ward1997} or high-eccentricity tidal migration \citep{Wu2003, Storch2014, Dawson2018, Jackson2021}.

As of April 23, 2024, a total of 6,438 exoplanets have been confirmed \footnote{http://exoplanet.eu/; http://exoplanets.org/; http://exoplanetarchive.ipac.caltech.edu/; http://openexoplanetcatalogue.com/; http://www.astro.keele.ac.uk/jkt/tepcat/rossiter.html.}. Among these, 4,861 are part of exoplanet systems, with 3,826 residing in single-star systems. Hot Jupiters are a class of gas giants with masses between 0.3 $M_{\rm J}$ and 13 $M_{\rm J}$ (where $M_{\rm J}$ is the mass of Jupiter), and orbital periods typically shorter than 10 days. Of the exoplanets discovered, 659 are confirmed as hot Jupiters, most of which exist alone without close companion planets. Only 44 hot Jupiters have companion planets, and all 99 hot Jupiters found in binary systems are classified as S-type. The primary detection method for observed hot Jupiters with companions is the transit method, which has strong selection biases favoring close-in, large planets, as well as multiplanet systems that are well-aligned. Data on projected obliquities is available for about 129 hot Jupiters, with roughly $90\%$ of them showing obliquities within the range of $[-149^\circ, 100^\circ]$.
%

Most lonely hot Jupiters have been found to have highly inclined or even retrograde orbits relative to the spin of their host stars \citep{Hebrard2008, Hebrard2010, Narita2009, Winn2009, Triaud2010, Albrecht2012, Winn2015}. These planets are typically found in exoplanet systems with high-mass, hot stars \citep{WangXY2022}, while cooler stars generally host exoplanets with lower obliquities \citep{Winn2010, Schlaufman2010, Albrecht2012}. \cite{Winn2010} proposed that all hot Jupiters initially possess high obliquities, which are then damped primarily by cool host stars. Under this assumption, hot Jupiters should rarely enter into mean-motion resonance with other planets. However, the case of WASP-5 b, whose transit timing variations suggest the presence of companions \citep{Fukui2011}, raises the possibility of alternative formation channels for hot Jupiters.

\cite{Batygin2012} pointed out that the torque exerted by distant giant companion stars could lead to the tilt of the gaseous disk, allowing hot Jupiters with native obliquities to form directly within the tilted disk. Additionally, magnetic interactions between stars and planetary disks may cause tilts in either the spin axis of the planetary disk or the host star \citep{Lai2011, Foucart2011}. The planetary disk may also experience tilting due to chaotic accretion in the later stage of star formation \citep{Bate2010}. Furthermore, the transport of angular momentum by internal gravitational waves in hot stars \citep{Rogers2012, Rogers2013} could induce a non-zero angle between the planetary disk and the host star’s spin axis. After planet formation, gravitational interactions between celestial bodies can also lead to high obliquities. Many mechanisms have been proposed to explain the high obliquities of planets, including Kozai cycles \citep{Wu2003, Fabrycky2007, Naoz2011, Wang2017}, planet-planet scattering \citep{Rasio1996, Ford2008, Nagasawa2008, Beauge2011, Nagasawa2011}, and secular chaos \citep{Wu2011, Batygin2012}.

With the continuous expansion of celestial mechanics and the increasing need for high-precision measurements, Einstein's general theory of relativity, proposed in 1916 \citep{Iorio2015, Debono2016, Vishwakarma2016, Beltran2019}, has become an important area of study. The discovery of many massive, close exoplanets has increased the possibility of testing general relativity in planetary systems \citep{Brumberg1991, Soffel2019, Iorio2024}. Among the general relativistic effects that may be detected are first-order effects, such as the general relativistic gravitoelectric correction to the Keplerian orbital period \citep{Iorio2006}, and general relativistic precession of periastron \citep{Jordan2008, Jordan2009}, as well as second-order effects such as the precession of the orbital node \citep{Iorio2011a, Iorio2022b, Iorio2024} and spin effects \citep{Iorio2021, Iorio2022a} due to de Sitter and Lense-Thirring effects. However, we should exclude precession effects induced by the gravitational quadrupole fields created by theirself rotation, tidal bulges and other disturbing objects \citep{Ragozzine2009, Kane2012}. Exoplanet systems offer valuable opportunities to test fundamental gravitational theories, modified theories of gravity, and alternative relativistic theories through observations of secular Transit Timing Variations (TTV) \citep{Li2012, Zhao2013} and the Radial Velocity method \citep{Gallerati2022}.

\cite{Iorio2011b} discussed the general relativistic impacts on directly observable quantities in transiting exoplanets, such as transit duration, radial velocity, the time interval between primary and secondary eclipses, and the period between successive primary transits. Transit duration variations are considered one of the most promising methods \citep{Eibe2012} to detect general relativistic precession of periastron over an observation period of about 4 years \citep{Pal2008}. This method also provides an alternative approach to constraining obliquity, besides the Rossiter-McLaughlin effect \citep{Damiani2011}. By accurately measuring the time elapsed between primary mid-transit and secondary mid-eclipse, \cite{Blanchet2019} proposed a method for detecting relativistic effects in the orbital motion of high-eccentricity exoplanets such as HD 80606b and WASP-14 b \citep{Antoniciello2021}. However, secondary transits are generally weak and difficult to observe, making the Radial Velocity method a more feasible alternative \citep{Gou2021}.

Under the framework of general relativity, the motion of most celestial bodies in the weak gravitational field of the solar system is generally described using the post-Newtonian approximation, which is divided into the post-Newtonian Lagrange approximation in harmonic coordinates \citep{Kidder1995, Andrade2001, Faye2006, Chen2016} and the post-Newtonian Hamiltonian approximation in ADM coordinates \citep{Damour2001a, Damour2001b, Faye2006, Jaranowski2015}. For hot Jupiter systems, the magnitude of the first-order post-Newtonian term relative to the Newtonian term is on the order of $10^{-7}$, meaning that relativistic effects should be considered in long-term dynamical evolution. \cite{Biscani2013, Biscani2015} focused on the long-term evolution of the spin and orbit of the secondary celestial body, using modern perturbation methods based on Lie series to analyze relativistic spin-spin and spin-orbit interactions. In their study, the spin of the host star was treated as a constant of motion in the restricted case and was excluded from the Hamiltonian. However, our research shows that relativistic effects from a close planet can accumulate over time, inducing substantial changes in the spin of the host star, which could lead to precession of the star’s spin axis and, consequently, an increase in obliquity—the angle between the star’s spin and the orbital plane of other planets or the planetary disk. In this paper, we primarily focus on the 1.5-order post-Newtonian two-body problem with spins to analyze the influence of the spin-orbit coupling term on the precession of both the host star’s and the planet’s spins.

The rest of this paper is organized as follows: In Section \ref{ch.model}, we introduce the Hamiltonian equations for the first-order post-Newtonian two-body problem with spins, transforming the non-canonical spin variables to obtain the Hamiltonian equations with a global symplectic structure. Section \ref{ch.integratility} discusses the integrability of the Hamiltonian describing the first-order post-Newtonian two-body problem with orbit-spin coupling. In Section \ref{ch.evolution}, we present numerical simulations of hot Jupiter systems and analyze the influence of the spin-orbit coupling post-Newtonian term on the host star’s spin for different initial inclinations of the hot Jupiter. In Section \ref{ch.spinspin}, the influence of 2PN spin-spin coupling term on the periods of spin precession are discussed. Finally, the conclusion are provided in Section \ref{ch.con}. 

\section{Hamiltonian formulation}
\label{ch.model}
This paper adopts the 1.5-order post-Newtonian approximation for the two-body problem with spins as the model of choice. After transforming to the center-of-mass frame, the Hamiltonian is defined according to \citep{Barker1970, Barker1979, Damour2001a}.
\begin{eqnarray}
H=H_{\rm N}+\epsilon H_1.
\label{eq1}
\end{eqnarray}
Here $\epsilon  =1/c^2$, c is the speed of light.\\
$H_1$ expands as
\begin{eqnarray}
H_1=H_{\rm 1PN}+H_{\rm SO}.
\label{eq2}
\end{eqnarray}
Their detailed expressions are defined by
\begin{eqnarray}
H_{\rm N}=\frac{1}{2}\frac{\boldsymbol{J}_1^2}{I_1}+\frac{1}{2}\frac{\boldsymbol{J}_2^2}{I_2}+\frac{\mathbf{\boldsymbol{p}}^2}{2{\mu}}-\frac{GM\mu}{r},
\label{eq3}
\end{eqnarray}
which is the Newtonian Hamiltonian,

\begin{eqnarray}
H_{\rm 1PN}&=&\mu \{ \frac{1}{8} (3\nu -1)\frac{\mathbf{\boldsymbol{p}} ^4}{ \mu ^4}  \nonumber\\&&
-\frac{GM}{2r} [ (3+\nu )\frac{\mathbf{\boldsymbol{p}} ^2}{\mu ^2}
+\nu ( \mathbf{\boldsymbol{n}}\cdot \frac{\mathbf{\boldsymbol{p}}}{\mu }) ^2] +
\frac{G^2M^2}{2r^2}  \},
\label{eq4}
\end{eqnarray}
which is the 1PN orbital Hamiltonian, and 

\begin{eqnarray}
H_{\rm SO}&=&\frac{2G}{r^3} [ ( 1+\frac{3}{4} \frac{m_2}{m_1})  \boldsymbol{J} _1  \nonumber\\&&
+( 1+\frac{3}{4}\frac{m_1}{m_2} )\boldsymbol{J} _2] \cdot(  \boldsymbol{r}\times  \boldsymbol{p} ),
\label{eq5}
\end{eqnarray}
which is the 1.5PN spin-orbit part.
G is the gravitational constant,
$m_1$, and $m_2$ are the mass of the secondary and main celestial bodies respectively.
$M=m_1+m_2$ is the total mass, while $\mu =m_1m_2/M $ is the reduced mass, and $\nu  =\mu / M $.
$\boldsymbol{p}=\boldsymbol{p}_1=\boldsymbol{-p}_2$,
$\boldsymbol{p}_1$ and $\boldsymbol{p}_2$ are momenta of body 1
and 2, respectively, in the center-of-mass frame.
$\boldsymbol{r}=r \boldsymbol{n}$ is the vector connecting primary celestial
body 2 and secondary celestial body 1. $\boldsymbol{n}$ denotes the unit vector in the direction of the separation between the two bodies.

If the spin vectors $\boldsymbol{J}_i$ in Hamiltonians
of equations \ref{eq3} and \ref{eq5} are expressed using the general form of non-canonical spin variables,
it may hinder the deduction of the integrability and the further application of the symplectic algorithm in numerical simulation which is very suitable for long-term dynamical evolution. \citet{Wu2010} proposed cylindrical-like spin coordinates to represent spin vectors.
Here, we transform the spin vector $\boldsymbol{J}_i$ into the canonical coordinates $\boldsymbol{\theta}_i$
and canonical momenta $\boldsymbol{\xi}_i$. Therefore, the canonical spin variable can be expressed as

\begin{eqnarray}
  \boldsymbol{J}_i=\left( \begin{array}{ccc} \sqrt{J_i^2-\xi _i^2}\cos \theta _i \\\sqrt{J_i^2-\xi _i^2}\sin  \theta _i\\\xi _i^2
\end{array}\right) (i=1,2).
\label{eq6}
\end{eqnarray}

$\boldsymbol{J}_i$ is regarded as the rotational angular momentum of spherical
rigid bodies \citep{Barker1975, Barker1976, Barker1979, Wex1995, Biscani2013, Biscani2015}, which can be expressed as
\begin{equation}
  \boldsymbol{J}_i =I_i \boldsymbol{\omega}_i,
\end{equation}
where $ \boldsymbol{\omega}_i $ is the rotational angular
velocity vector of body \emph{i} and  $I_i $ is the moment of inertia.
The moment of inertia is related to the mass and radius of
celestial bodies and is expressed as
\begin{equation}
  I_i\approx \alpha mR^2.
\end{equation}
When $I_i $ represents the moment of inertia of the stars,
the value of $\alpha$ ranges from approximately 0.031 to 0.148 for main sequence stars \citep{Claret1989} .
When $I_i $ refers to the moment of inertia of planets,
 $\alpha$ is approximately 0.2 for gas giants and 0.25 for solid planets
\citep{Ward2004,Millholland2018}.

In the center-of-mass coordinate system, the final simplified Hamiltonian expressed in terms of canonical variables
($\boldsymbol{r}, \boldsymbol{p}, \boldsymbol{J}_1, \boldsymbol{J}_2$)
 can be written as
\begin{eqnarray}
  H(\mathbf{r}, \boldsymbol{p}, \boldsymbol{J}_1, \boldsymbol{J}_2 )
  = H(\mathbf{r}, \boldsymbol{p}, \boldsymbol{J}_1(\theta_1, \xi _1), \boldsymbol{J}_2(\theta_2, \xi _2)),
\label{eq9}
  \end{eqnarray}
where
\begin{eqnarray}
H_{\rm SO}&=& \frac{2G}{r^3} \{ [ (1+\frac{3}{4}\frac{m_2}{m_1} ) (\sqrt{J_1^2-\xi _1^2}\cos \theta _1 )\nonumber\\&&
+(1+\frac{3}{4}\frac{m_1}{m_2} ) (\sqrt{J_2^2-\xi _2^2}\cos \theta _2 )] (yp_x-zp_y)\nonumber\\&&
[ (1+\frac{3}{4}\frac{m_2}{m_1} ) (\sqrt{J_1^2-\xi _1^2}\sin  \theta _1 )+(1+\frac{3}{4}\frac{m_1}{m_2} )\nonumber\\&&
(\sqrt{J_2^2-\xi _2^2}\sin \theta _2 )] (zp_x-xp_z) +[ (1+\frac{3}{4}\frac{m_2}{m_1} ) \nonumber\\&&\xi _1
+(1+\frac{3}{4}\frac{m_1}{m_2} )\xi _2](xp_y-yp_x)\}.
\end{eqnarray}

\section{The integrability of the system}
\label{ch.integratility}
For a general post-Newtonian double-spinning two-body problem, the position $\boldsymbol{r}$ and momentum $\boldsymbol{p}$ are a set of canonical variables that satisfy the Hamiltonian equations of motion,

\begin{equation}
\frac{{d\boldsymbol{r}}}{{dt}} = \frac{{\partial H}}{{\partial \boldsymbol{p}}},\frac{{d\boldsymbol{p}}}{{dt}} =  - \frac{{\partial H}}{{\partial \boldsymbol{r}}}.
\label{eq16}
\end{equation}

The spin-evolution equations are
\begin{equation}
\frac{{d{\boldsymbol{J}_i}}}{{dt}} = \frac{{\partial {H_{SO}}}}{{\partial {\boldsymbol{J}_i}}} \times {\boldsymbol{J}_i}.
\label{eq17}
\end{equation}
 The existence of six integrals of motion in the 12-dimensional phase space formed by $\left[ \boldsymbol{r}, \boldsymbol{p}, \boldsymbol{J}_1, \boldsymbol{J}_2 \right]$, —including the total energy $H$, the constant magnitudes of spins $J_{1}, J_{2}$, and the total angular momentum $\boldsymbol{J} = \boldsymbol{L}+ \boldsymbol{J}_1 + \boldsymbol{J}_2$ ( $\boldsymbol{L}=\boldsymbol{r}\times \boldsymbol{p}$ is the orbital angular momentum) —does not necessarily imply that the Hamiltonian is integrable, according to Liouville's theorem. This theorem states that a canonical Hamiltonian with $n$ degrees of freedom is integrable if and only if there are $n$ independent isolating integrals\citep{Lichtenberg(2013)}. Thus, the integrability of canonical equation \ref{eq16}  needs 3independent isolating integrals, while the integrability of non-canonical equation \ref{eq17}  needs 6 independent isolating integrals. In total, 9 independent isolating integrals are needed to confirm the integrability of the Hamiltonian described by equation\ref{eq5}. 

After dopting  the canonical spin variables in equation \ref{eq6},  each 3-dimensional variable  $\boldsymbol{J}_i$ is reduced to  2 degrees of freedom, with the condition that the magnitude of the vector $\boldsymbol{J}_i$  remains constant. The canonical spin Hamiltonian equations can then be derived as follows:
\begin{equation}
\frac{{d{\theta _i}}}{{dt}} = \frac{{\partial {H_{SO}}}}{{\partial {\xi _i}}},\frac{{d{\xi _i}}}{{dt}} =  - \frac{{\partial {H_{SO}}}}{{\partial {\theta _i}}}.
\label{eq18}
\end{equation}

The Hamiltonian $H(\mathbf{r}, \boldsymbol{p}, \boldsymbol{J}_1(\theta_1, \xi _1), \boldsymbol{J}_2(\theta_2, \xi _2))$ consists of 10-dimensional variables with  5 degrees of freedom. Its integrability needs 5 independent isolating integrals.  Besides the total energy $H$ and the total angular momentum $\boldsymbol{J}$, \citet{Damour2001a} gave out that post-Newtonian Hamiltonian containing  orbital PN orders and the 1.5PN spin-orbit coupling has two additional conserved quantities
\begin{equation}  
\boldsymbol{L} \cdot \boldsymbol{L} = const, \boldsymbol{L} \cdot {\boldsymbol{S}_{eff}} = const,
\label{con}
\end{equation}
where, ${\boldsymbol{S}_{eff}} = [2 + 3{m_2}/({m_1})]{\boldsymbol{J}_1} + [2 + 3{m_1}/({m_2})]{\boldsymbol{J}_2}$.
The six integrals of motion can determine the integrability of the canonical Hamiltonian described by equations \ref{eq1}-\ref{eq4} and \ref{eq9}. Furthermore, since integrability require only five independent integrals, the additional integral implies that two of the five frequencies are commensurable. Thus, the integrability of the system means that there are no chaotic orbits, only periodic and quasi-periodic orbits.

\section{The dynamical evolutions of the two spinning bodies considering 1.5 PN term}
\label{ch.evolution}

As is well known, post-Newtonian terms introduce relativistic effects that apply to the motion of celestial bodies in weak gravitational fields and at low velocities. This implies that the orbital angular momentum remains approximately constant when considering post-Newtonian corrections, whereas it is strictly conserved in two-body problems that only consider Newtonian dynamics.

\begin{figure}
\epsscale{1.0}
\centering
\includegraphics[width=4in,height=3.4in]{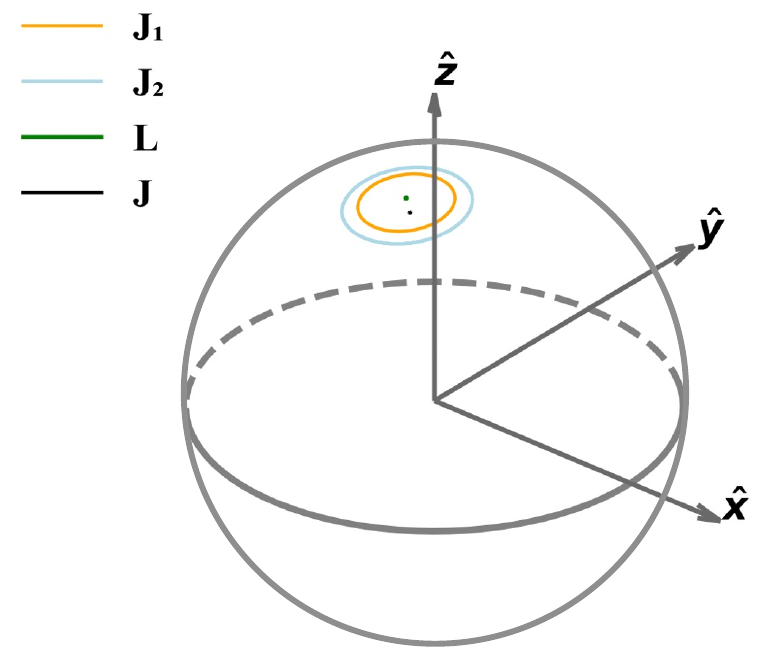}
\caption{The projections of the spins $\mathbf{J}_1$, $\mathbf{J}_2$, the orbital angular momentum $\mathbf{L}$, and the constant $\mathbf{J}$ onto a 3-dimensional spherical surface are shown. The initial orbital parameters, with eccentricity $e = 0.5$, are listed in Case-0 of Table \ref{Table2}.}
\label{projection}
\end{figure}

\begin{figure}
\epsscale{1.0}
\centering
\plotone{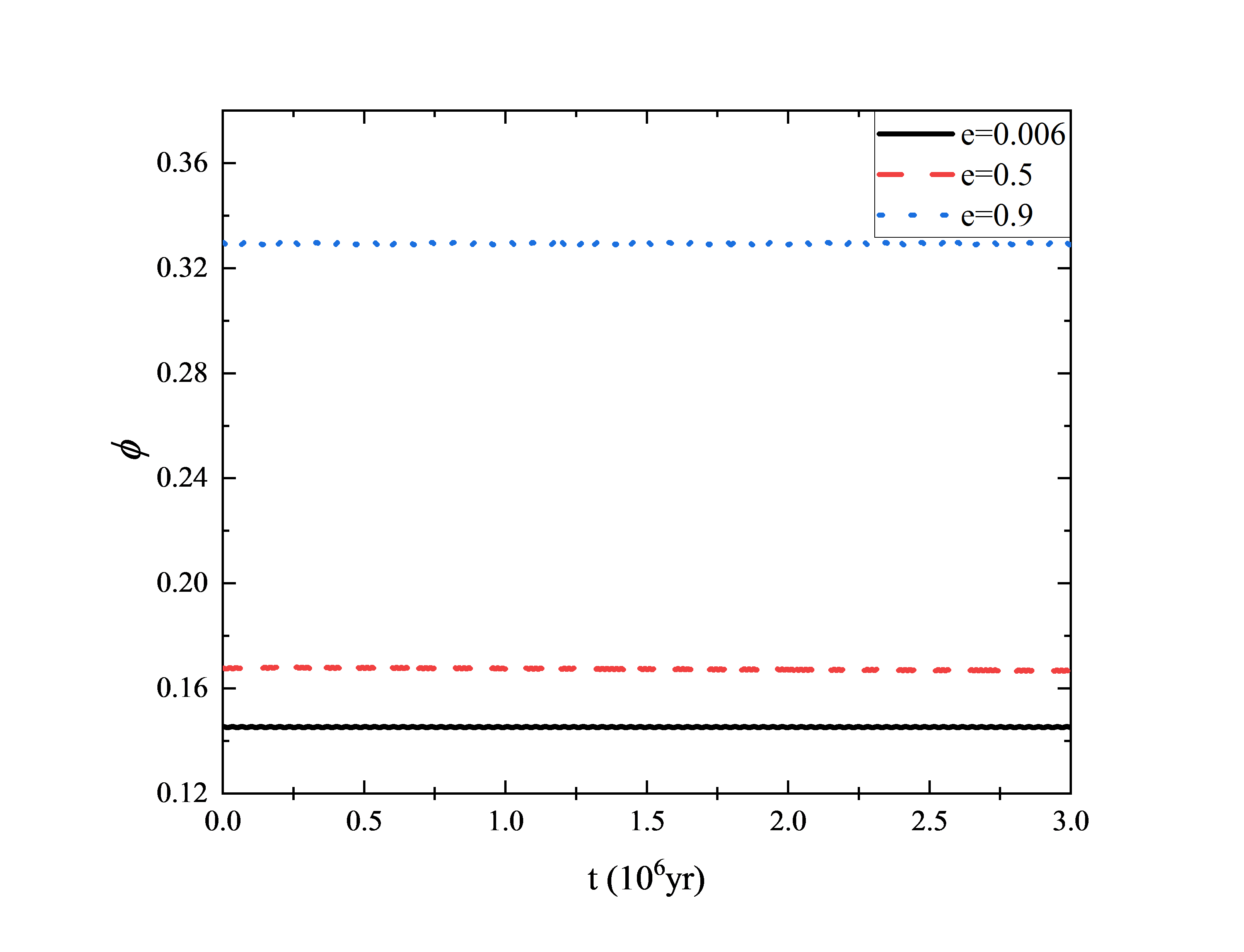}
\caption{The numerical evolutions of the angle between the total angular momentum $\mathbf{J}$ and the orbital angular momentum $\mathbf{L}$ for three different eccentricities, $e = 0.006$, $0.5$, and $0.9$, are shown. The other initial orbital parameters are provided in Case-0 of Table \ref{Table2}.}
\label{LJ}
\end{figure}

We plot the projections of the spins $\mathbf{J}_1,\mathbf{J}_2$, the orbital angular momentum $\mathbf{L}$, and the constant $\mathbf{J}$ on a 3-dimensional spherical surface, as shown in Figure.\ref{projection}. Case-0 in Table.\ref{Table2} provides the initial orbital parameters of an orbit with a mild eccentricity $e=0.5$. As shown, $\mathbf{L}$ exhibits only a slight variation, remaining nearly constant, while $\mathbf{J}$ keeps constant strictly. Additionally, in Figure.\ref{LJ},  the variations in the angles between $\mathbf{L}$ and $\mathbf{J}$  are almost imperceptible for different eccentricities ($e=0.006,0.5,0.9$).

\begin{table}
\begin{center}
\caption{Summary and explanation of the symbolic variables appearing in this article.}
\label{Table1}
\begin{tabular}{cccc}
\hline
\textbf{ Parameter}&\textbf{unit}&\textbf{Meaning}\\
\hline
$M_1$ & $M_ J$      & Mass of planet \\
$M_2$ & $M_\odot $  & Mass of primary\\
$R_1$ & $R_J $ & Radius of planet\\
$R_2$ & $R_\odot $ & Radius of primary\\
$a$ & $au $         & Orbital semi-major axis\\
$e$ & $\backslash$  & Orbital eccentricity\\
$i$ & $deg$         & Orbital inclination\\
$\Omega$& degree    & longitude of ascending node\\
$\omega$& degree    &argument of pericenter\\
$M$ & degree        &mean anomaly\\
$P_{2rot}$ &day     & Rotation period of primary\\
$P_{1rot}$ &day     & Rotation period of planet\\
$\theta_1$ & degree &Nodal angle of planetary spin\\
$\theta_2$ & degree &Nodal angle of star spin \\
$\varphi_1$ & degree   & Angle between $\mathbf{J}_1$and Z axis\\
$\varphi_2$ & degree   & Angle between $\mathbf{J}_2$ and Z axis\\
$\lambda_1 $ & degree   & Angle between $\mathbf{J}_1$ and $\mathbf{L}$\\
$\lambda_2 $ & degree   & Angle between $\mathbf{J}_2$ and $\mathbf{L}$\\
$\phi $ & degree   & Angle between $\mathbf{J}$ and $\mathbf{L}$\\
\hline
\end{tabular}
\end{center}
\end{table}

\begin{table}
\begin{center}
\caption{The initial parameters for the three numerical cases are provided, with their units and explanations listed in Table \ref{Table1}.}
\label{Table2}
\begin{tabular}{cccc}
\hline
\textbf{Parameter}&\textbf{Case-0}&\textbf{Case-1}&\textbf{Case-2} \\
\hline
$M_1$ & 10   & 1     & 10 \\
$M_2$ & 1.09& 1.2  & 1\\
$R_1$ &  1     &  1    & 1\\
$R_2$ &  1     &  1    & 1\\
$a$     &  0.05  &0.04 & 0.05\\
$e$   &~&  0.004& 0.006\\
$i$   &10&  30   & [0,20]\\
$\Omega$&0& 0   & 0\\
$\omega$& 0&0   & 0\\
$M$ &   0&  0   & 0\\
$P_{2rot}$ &30&25& 30\\
$P_{1rot}$ &4&4 & 4\\
$\theta_1$ &90&90& 90\\
$\theta_2$ &95&95& 95 \\
$\varphi_1$ &1 &1& 1\\
$\varphi_2$ &5&1& 5\\
\hline
\end{tabular}
\end{center}
\end{table}

\begin{figure}
\epsscale{1.0}
\centering
\includegraphics[width=4in,height=3.4in]{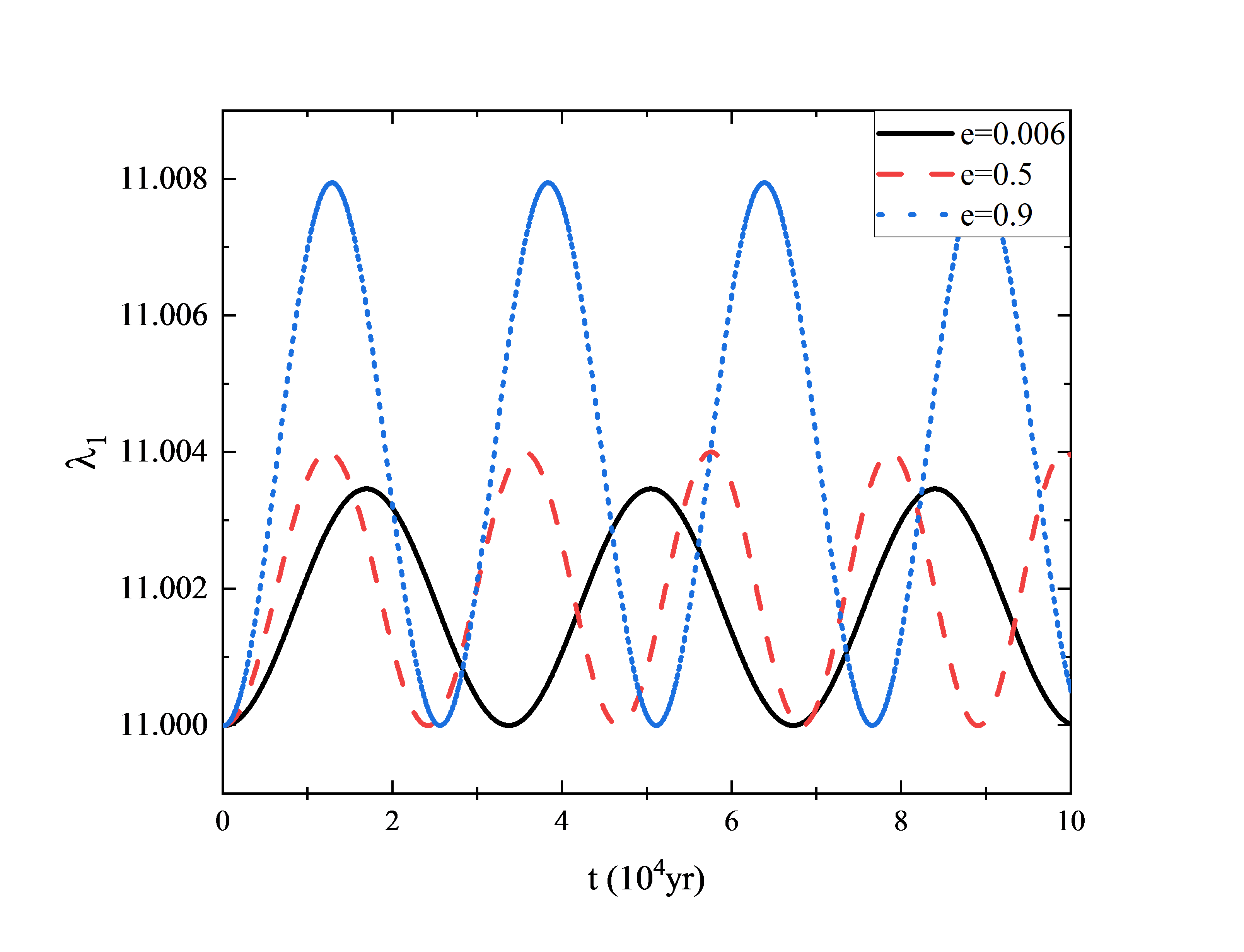}
\includegraphics[width=4in,height=3.4in]{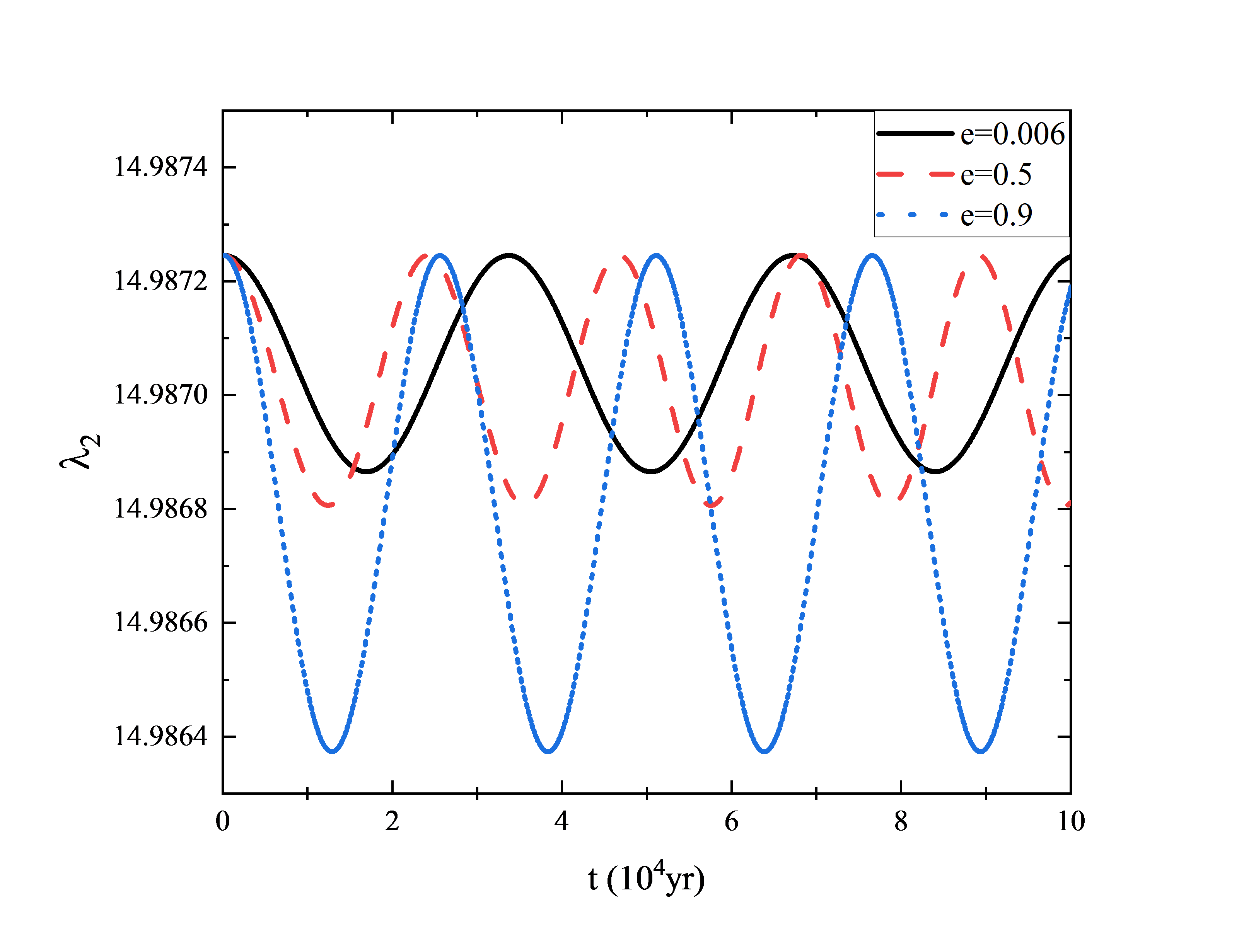}
\caption{The numerical evolutions of the angles between the spins $\mathbf{J}_1$, $\mathbf{J}_2$ and the orbital angular momentum $\mathbf{L}$ for three different eccentricities, $e = 0.006$, $0.5$, and $0.9$, are shown. (a) The angle between $\mathbf{J}_1$ and $\mathbf{L}$, and (b) the angle between $\mathbf{J}_2$ and $\mathbf{L}$. The other initial orbital parameters are provided in Case-0 of Table \ref{Table2}.}
\label{S12L}
\end{figure}


However, the spins rotate around $\mathbf{L}$ with the angles changing slightly and periodically, as shown in Figure \ref{S12L}. This implies that the obliquities (the angle between the spins and the orbital angular momentum) remain nearly constant. It is important to note that the obliquity often referred to in various studies is $\lambda_2$, the angle between the spin of the host star and the normal to the planetary orbital plane. While the 1.5PN spin-orbit coupling term does not significantly alter the obliquities in a system of two spinning bodies, it can cause the spatial orientations of the spins to change considerably.

\begin{figure}
\epsscale{0.7}
\centering
\includegraphics[width=4in,height=3.4in]{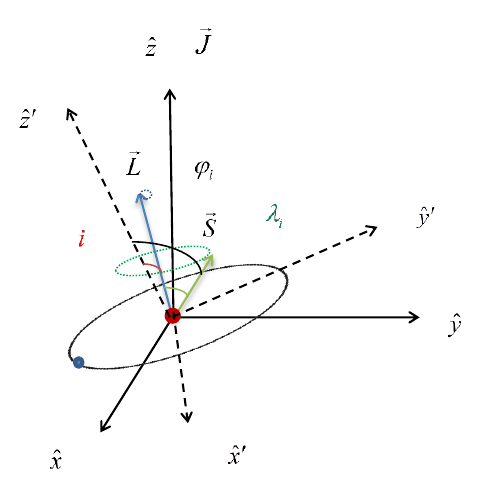}
\includegraphics[width=4in,height=3.4in]{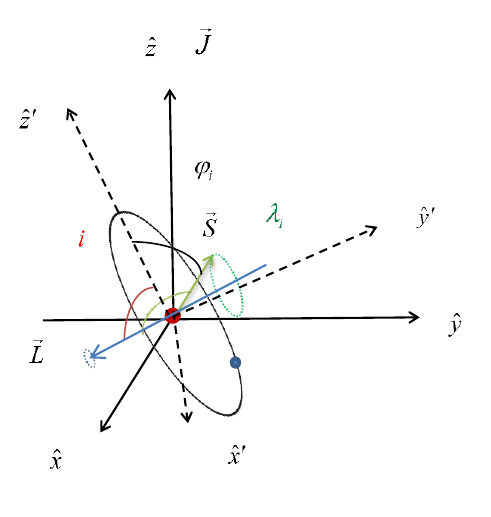}
\caption{$\mathord{\buildrel{\lower3pt\hbox{$\scriptscriptstyle\frown$}} \over x} \mathord{\buildrel{\lower3pt\hbox{$\scriptscriptstyle\frown$}} \over y} \mathord{\buildrel{\lower3pt\hbox{$\scriptscriptstyle\frown$}} 
\over z}$ coordinate system is the invariable plane coordinate system. $\mathord{\buildrel{\lower3pt\hbox{$\scriptscriptstyle\frown$}} 
\over x} '\mathord{\buildrel{\lower3pt\hbox{$\scriptscriptstyle\frown$}} 
\over y} '\mathord{\buildrel{\lower3pt\hbox{$\scriptscriptstyle\frown$}} 
\over z} '$ is the general coordinate system. $\mathord{\buildrel{\lower3pt\hbox{$\scriptscriptstyle\frown$}} 
\over z}$ axis can be set parallel to the normal of the planetary disk without generality. $\mathbf{J}$ keeps constant strictly. $\mathbf{L}$ keeps constant approximately. The spins $S$ rotate around $L$ periodicaly.
}
\label{precession}
\end{figure}

To vividly illustrate the revolutions of $\mathbf{L}$ and the spin $\mathbf{S}$, as well as the variation range of the spins, we show their loops in both the invariant plane coordinate system and the general coordinate system in Figure~\ref{precession}. The total angular momentum $\mathbf{J}$ is strictly conserved, while $\mathbf{L}$ remains approximately constant, with only a slight rotation. The spins $\mathbf{S}$ rotate periodically around $\mathbf{L}$.

From this, we can easily derive the rules for the spin deviation amplitude from the z-axis in the general coordinate system, which are determined by their respective obliquities $\lambda_i$ and the orbital inclination $i$. For example, the approximate maximum value of $\varphi_2$ (the angle between the spin of the host star and the z-axis) is $\lambda_2 + i$ for a prograde orbit, $2\pi - (\lambda_2 + i)$ for a retrograde orbit, and the approximate minimum value is $\left|\lambda_2 - i\right|$. Interestingly, the deviation between the host star's spin and the z-axis, $\varphi_2$, becomes significant for highly inclined orbits, even when the host star's spin is nearly parallel to the z-axis initially.

In this paper, all celestial bodies are assumed to have spherical symmetry, as shape asymmetry can also induce precession of the rotation axis. Therefore, the precession discussed here is solely attributed to the spin-orbit coupling post-Newtonian term. The amplitude of the angle between the spin orientation of the host star and the z-axis can be excited by an inclined planetary orbit. Without loss of generality, we define the normal direction of the planetary disk plane as the z-axis. Due to relativistic precession, the spin orientation of the host star and the normal direction of the planetary disk plane can deviate significantly, which may play a crucial role in the formation of exoplanet system obliquities. In the following section, we will explore in detail the impact of this relativistic dynamical mechanism on the variation of the spatial orientations of spins.

\subsection{The influence of inclined planetary orbits on relativistic spins precession: numerical simulations}
\label{ch.simulation}

In this section, we aim to examine the evolutionary characteristics of the spins due to the 1.5 PN orbit-spin coupling term. We simulate the evolution of the spins in a hot Jupiter system by incorporating the first-order post-Newtonian Hamiltonian and the 1.5 PN spin-orbit coupling term. Post-Newtonian Hamiltonian systems of spinning two-body problem have no separable form of coordinates and momenta or several integrable splitting parts \citep{Wang2021a,Wang2021b,Wang2021c, Sun2021,Wu2021}. Explicit symplectic integrators cann't be applied directly.  
We applied  the fourth-order extended phase space symplect-like method with midpoint permutations \citep{Luo2017a, Luo2017b}. We assume that the positive z-axis is aligned with the initial normal direction of the planetary disk. In the Solar system, the Sun's spin deviation from the normal of the ecliptic plane is about $7^\circ$, and the Earth's obliquity is about $23^\circ$. Based on these values, we set the initial angles $\varphi_1 = 1^\circ$ and $\varphi_2 = 5^\circ$, which are reasonable deviations from the normal direction of the planetary disk and smaller than those observed in the Solar system.

\begin{figure}
\epsscale{1.0}
\centering
\includegraphics[width=4in,height=3.4in]{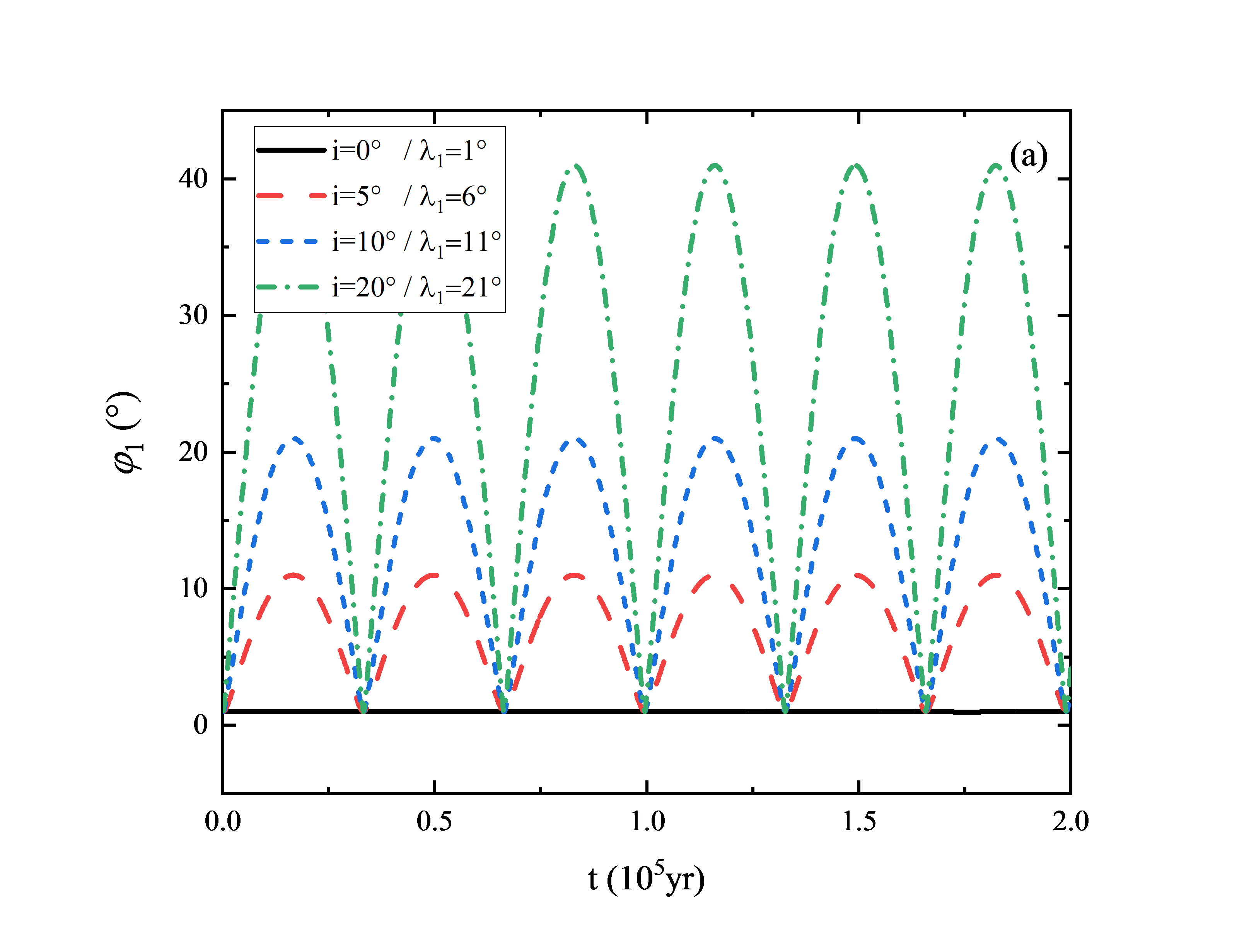}
\includegraphics[width=4in,height=3.4in]{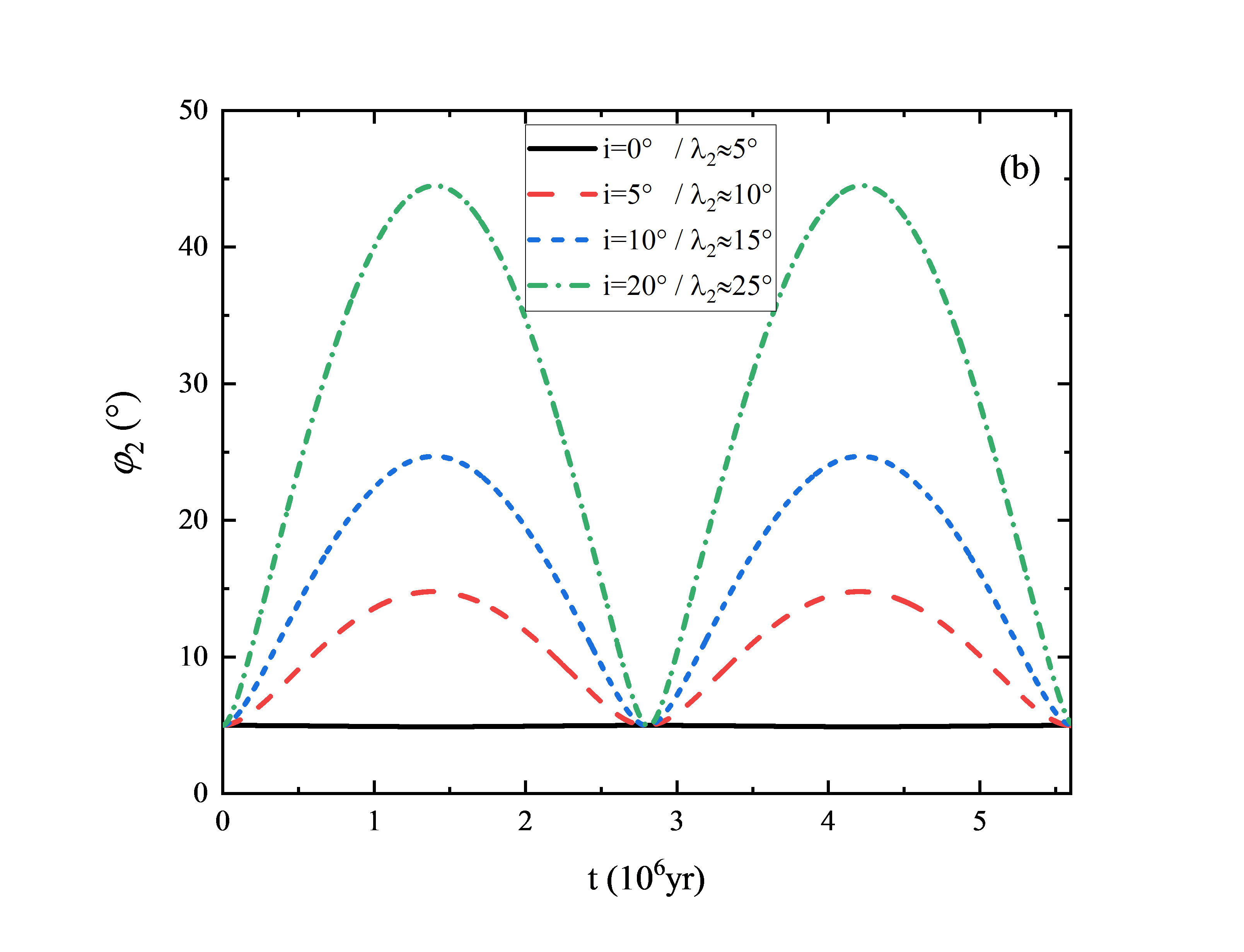}
\caption{The spin precession of the planet and host star in a system with inclined planetary orbits, induced by the 1.5-order orbital-spin coupling post-Newtonian term, is examined for inclination angles $i = 0^\circ$, $5^\circ$, $10^\circ$, and $20^\circ$. The corresponding initial obliquities, $\lambda_1$, are calculated to be $1^\circ$, $6^\circ$, $11^\circ$, and $21^\circ$, respectively, while the initial obliquities for the host star, $\lambda_2$, are approximately $5^\circ$, $10^\circ$, $15^\circ$, and $25^\circ$, respectively. The other initial parameters of the system are listed in Case-2 of Table \ref{Table2}. }
\label{orientation}
\end{figure}

Figure \ref{orientation} (a) and (b) show the time evolutions of the angles between the spins and the initial normal direction of the planetary disk, denoted as $\varphi_1$ and $\varphi_2$, respectively. It is evident that the maximum value of $\varphi_i$ is $\lambda_i + i$, and the minimum value is $|\lambda_i - i|$, which is consistent with the theoretical analysis presented in the previous section.

From the numerical results above, we can obtain the evolution period of $\lambda_1$ (the angle between the planet's spin and the normal to the planetary disk, or the orientation of the z-axis), which is $3.32 \times 10^4$ years. The precession period of $\lambda_2$ (the angle between the host star's spin and the normal to the planetary disk, or the orientation of the z-axis) is $2.81 \times 10^6$ years. Notably, the precession period is independent of the orbital inclination, as it is relative to any fixed plane. In the next section, we will derive the theoretical expression for the spin precession period.

\subsection{ Relativistic spins precession period: theoretical analysis in general case}
\label{ch.period}

We can derive the theoretical spin precession period of the planet using restricted models from previous literature. One such model is the well-known relativistic effect called geodetic or de Sitter precession, which describes the spin precession of a gyroscope orbiting a large (non-rotating) spherical mass 
$m_{2}$ in a circular orbit \citep{DeSitter1916, Barker1970, Barker1979, Schiff1960a, Schiff1960b}.

\begin{equation}
{{\mathord{\buildrel{\lower3pt\hbox{$\scriptscriptstyle\frown$}} 
\over T} }_1} = \frac{{4\pi {c^2}{a^{{5 \mathord{\left/
 {\vphantom {5 2}} \right.
 \kern-\nulldelimiterspace} 2}}}}}{{3{G^{{3 \mathord{\left/
 {\vphantom {3 2}} \right.
 \kern-\nulldelimiterspace} 2}}}m_2^{{3 \mathord{\left/
 {\vphantom {3 2}} \right.
 \kern-\nulldelimiterspace} 2}}}}
\label{eq.Period1}
\end{equation} 
where $\bar \omega  = {\left( {\frac{{G{m_2}}}{{{a^3}}}} \right)^{{1 \mathord{\left/{\vphantom {1 2}} \right.\kern-\nulldelimiterspace} 2}}}$.

The other model assumes that the ratio ${{J_1} \mathord{\left/ {\vphantom {J_1 J_2}} \right. \kern-\nulldelimiterspace} {J_2}}$ is sufficiently small (approximately ${10^{-3}} \sim {10^{-4}}$ for Jupiter-like planets). In this case, the spin precession of the planet can be derived under the restricted condition (${m_2} \gg {m_1}$ and $\left| {\boldsymbol{J}_2} \right| \gg \left| {\boldsymbol{J}_1} \right|$, where $\boldsymbol{J}_2$ is treated as a constant). A general theoretical analysis of the post-Newtonian two-body problem with spins under the restriction $\boldsymbol{J}_2 = \text{constant}$ has been thoroughly addressed in \citep{Biscani2013,Biscani2015}.

The precession period of spin $\boldsymbol{J}_1$ can be derived from  \citep{Biscani2013}
 \begin{equation}
{{\mathord{\buildrel{\lower3pt\hbox{$\scriptscriptstyle\smile$}} 
\over T} }_1} = \frac{{4\pi {c^2}{{\tilde L}^3}{{\tilde G}^3}}}{{3m_2^4{G^4}M}},
\label{eq.Period2}
\end{equation}  
where,$\tilde L = \sqrt {G{m_2}a}$, $\tilde G = \tilde L\sqrt {1 - {e^2}}$. $M$ is the magnitude of the total mean angular momentum vector in the scaled restricted post-Newtonian Hamiltonian, which can be obtained by 

 \begin{equation}
{M^2} = {\tilde G^2} - 2{\tilde H^2}_0 + 2{\tilde H_0}{\tilde J_{*}} + {\tilde J_1}^2 + 2{\tilde G_{xy,0}}{\tilde J_{1xy,0}}{\cosh _{ * ,0}},
\end{equation} 
where $\tilde H=\tilde G{ \cos i} $,  ${{{\tilde J}_ * } = \tilde H + {{\tilde J}_{1z}}}$, $\tilde J_{1}={{\left| {{{\bf{J}}_1}} \right|} \mathord{\left/{\vphantom {{\left| {{{\bf{J}}_1}} \right|} {{m_1}}}} \right.
 \kern-\nulldelimiterspace} {{m_1}}}$,${h_*} = \Omega  - {\theta _1}$. The subscript $_0$ , $_{xy}$, $_z$ represent the value of the initial time, the component of the xy plane, the z component of the value respectively.

When $e=0$ and $\boldsymbol{J}_2=0$ the Equation.\ref {eq.Period2} will be consistent with Equation.\ref {eq.Period1} of geodetic or de sitter precession.

The values calculated using Equation \ref{eq.Period1} and Equation \ref{eq.Period2} with the parameters from Case-2 are $3.34 \times 10^4$ years and $3.80 \times 10^4$ years, respectively, which closely approximate the numerical period of $3.32 \times 10^4$ years. It is important to note that these results are only approximate in the case of circular orbits within the restricted models. Furthermore, the theoretical spin precession period for the host star has not been derived yet. In the following section, we will provide the theoretical formula for the spin precession periods in a general two-body system with spins, considering the 1.5 post-Newtonian terms.


%
%

According to the discussion in Section \ref{ch.integratility}, the system consisting of two spinning celestial bodies at 1.5 post-Newtonian order is integrable. While an integrable system may not necessarily have explicit analytical solutions, theoretical analysis can provide more insights compared to non-integrable systems.

Equation~\ref{eq16} and \ref{eq18} represent the Hamiltonian vector field of the Hamiltonian $H$ from Equation~\ref{eq9}, obtained by calculating the Poisson brackets 
$\left\{ \sim , H \right\}$. Similarly, by following the flow of the generators $J^2$, $J_z$, and $L^2$, we can derive the equations for $\mathbf{r}$, $\boldsymbol{p}$, 
$\boldsymbol{J}_1$, and $\boldsymbol{J}_2$, along with their evolutionary behaviors, by computing the Poisson brackets $\left\{ { \sim , J^2} \right\}$, $\left\{ { \sim , J_z} \right\}$, and $\left\{ { \sim , L^2} \right\}$.

For $\mathbf{r}, \boldsymbol{p}$, they satisfy 
\begin{equation}
\frac{{d\mathbf{V}}}{{d{\lambda _1}}} = \left\{ {\mathbf{V},{J^2}} \right\} = 2\mathbf{J }\times \mathbf{V},
\end{equation}
\begin{equation}
\frac{{d\mathbf{V}}}{{d{\lambda _2}}} = \left\{ {\mathbf{V},{J_z}} \right\} = \hat z \times \mathbf{V},
\end{equation}
and
\begin{equation}
\frac{{d\mathbf{V}}}{{d{\lambda _3}}} = \left\{ {\mathbf{V},{L^2}} \right\} = 2\mathbf{L}\times \mathbf{V},
\end{equation}
with $\mathbf{V}$ representing any of $\mathbf{r}$ and $\boldsymbol{p}$.
The three equations make $\mathbf{r}$, and $\boldsymbol{p}$ rigidly rotate about the constant $\mathbf{J }$, the 
$\hat z$ axis, and the orbital angular momentum $\mathbf{L }$ in the loops generated by the flow $J^2$,$J_z$, and $L^2$.  According to the definition of $\mathbf{L }=\mathbf{r} \times \mathbf{p}$, $\mathbf{L }$ also rotates about the constant $\mathbf{J }$, the $\hat z$ axis, and itself in the loops respectively.

For $\boldsymbol{J}_1$ and $\boldsymbol{J}_2$, they satisfy 
\begin{equation}
\frac{{d\mathbf{S}}}{{d{\lambda _1}}} = \left\{ {\mathbf{S},{J^2}} \right\} = 2\mathbf{J }\times \mathbf{S},
\end{equation}
\begin{equation}
\frac{{d\mathbf{S}}}{{d{\lambda _2}}} = \left\{ {\mathbf{S},{J_z}} \right\} = \hat z \times \mathbf{S},
\end{equation}
and
\begin{equation}
\frac{{d\mathbf{S}}}{{d{\lambda _3}}} = \left\{ {\mathbf{V},{L^2}} \right\} =0,
\end{equation}
with $\mathbf{S}$ representing any of $\boldsymbol{J}_1$ and $\boldsymbol{J}_2$.
The three equations make spin $\mathbf{S}$ rigidly rotates about the constant $\mathbf{J }$ and the 
$\hat z$ axis in the loops generated by the flow $J^2$ and $J_z$, but fixed (not rotating about $\mathbf{L }$ in the loops generated by the flow $L^2$.

Although the evolutions of $\mathbf{r}$, $\boldsymbol{p}$, $\boldsymbol{J}_1$, and $\boldsymbol{J}_2$ in the Hamiltonian flow generated by $\left\{ { \sim , H} \right\}$ will differ from their respective loops generated by the flow $\left\{ { \sim , L^2} \right\}$, we can more easily calculate the periods of the loops around the $\hat{z}$ axis, $\mathbf{J}$, and $\mathbf{L}$ through theoretical analysis. For instance, the period $T_2$ of $\boldsymbol{J}_2$ around the $\hat{z}$ axis can be obtained by

\begin{eqnarray}
\begin{array}{l}
{T_i} = \frac{{2\pi }}{{\tilde {\omega_i} }},\\
\tilde {\omega_i}  =\left\langle {\frac{{d{\lambda _2}}}{{dt}}} \right\rangle  = \left\langle {{{\frac{{d{\mathbf{J}_i}}}{{dt}}} \mathord{\left/
 {\vphantom {{\frac{{d{\mathbf{J}_i}}}{{dt}}} {\frac{{d{\mathbf{J}_i}}}{{d{\lambda _2}}}}}} \right.
 \kern-\nulldelimiterspace} {\frac{{d{\mathbf{J}_i}}}{{d{\lambda _2}}}}}} \right\rangle .
\end{array}
\end{eqnarray}
\label{pofs}
for an entier loop of the spin in the flow generator $J_z$ requires the angle $\lambda_2$ to change by $2\pi$. 

\begin{equation}
\frac{{d{\boldsymbol{J}_1}}}{{dt}} =\frac{{2G}}{{{c^2}{r^3}}}\left( {1 + \frac{3}{4}\frac{{{m_2}}}{{{m_1}}}} \right)\mathbf{L} \times {\mathbf{J}_1}.
\label{st1}
\end{equation}

\begin{equation}
\frac{{d{\boldsymbol{J}_2}}}{{dt}} =\frac{{2G}}{{{c^2}{r^3}}}\left( {1 + \frac{3}{4}\frac{{{m_1}}}{{{m_2}}}} \right)\mathbf{L} \times {\mathbf{J}_2}.
\label{st2}
\end{equation}

The angle brackets denote averaging over a full period of spin precession. It is important to note that the division of vectors assumes they are parallel. Therefore, when calculating Equation~\ref{pofs}, we must transform the general coordinate system to one where the direction of $\mathbf{L}$ is aligned with the $z$-axis. Furthermore, considering the conservation of $L$, $J_1$, and $J_2$, the angle between the spin and the orbital angular momentum only varies slightly. Thus, the average of $\frac{1}{{r^3}}$ is taken over a period of mean motion,

\begin{equation}
\left\langle {\frac{1}{{{r^3}}}} \right\rangle  = \frac{1}{{{a^3}{{\left( {1 - {e^2}} \right)}^{{3 \mathord{\left/
 {\vphantom {3 2}} \right.
 \kern-\nulldelimiterspace} 2}}}}},
\end{equation}
the varying period of $\mathbf{J_1}$ and $\mathbf{J_2}$ around $\hat z$ can be obtained approximatively by

\begin{equation}
{T_1} \approx \frac{{2\pi {c^2}{a^{{5 \mathord{\left/
 {\vphantom {5 2}} \right.
 \kern-\nulldelimiterspace} 2}}}\left( {1 - {e^2}} \right)\sqrt {{m_1} + {m_2}} }}{{{G^{{3 \mathord{\left/
 {\vphantom {3 2}} \right.
 \kern-\nulldelimiterspace} 2}}}\left( {2{m_1}{m_2} + 1.5{m_2}^2} \right)}}
\label{avT1}
\end{equation}

\begin{equation}
{T_2} \approx \frac{{2\pi {c^2}{a^{{5 \mathord{\left/
 {\vphantom {5 2}} \right.
 \kern-\nulldelimiterspace} 2}}}\left( {1 - {e^2}} \right)\sqrt {{m_1} + {m_2}} }}{{{G^{{3 \mathord{\left/
 {\vphantom {3 2}} \right.
 \kern-\nulldelimiterspace} 2}}}\left( {2{m_1}{m_2} + 1.5{m_1}^2} \right)}}
\label{avT2}
\end{equation}

\begin{figure}
\epsscale{1.0}
\centering
\includegraphics[width=4in,height=3.4in]{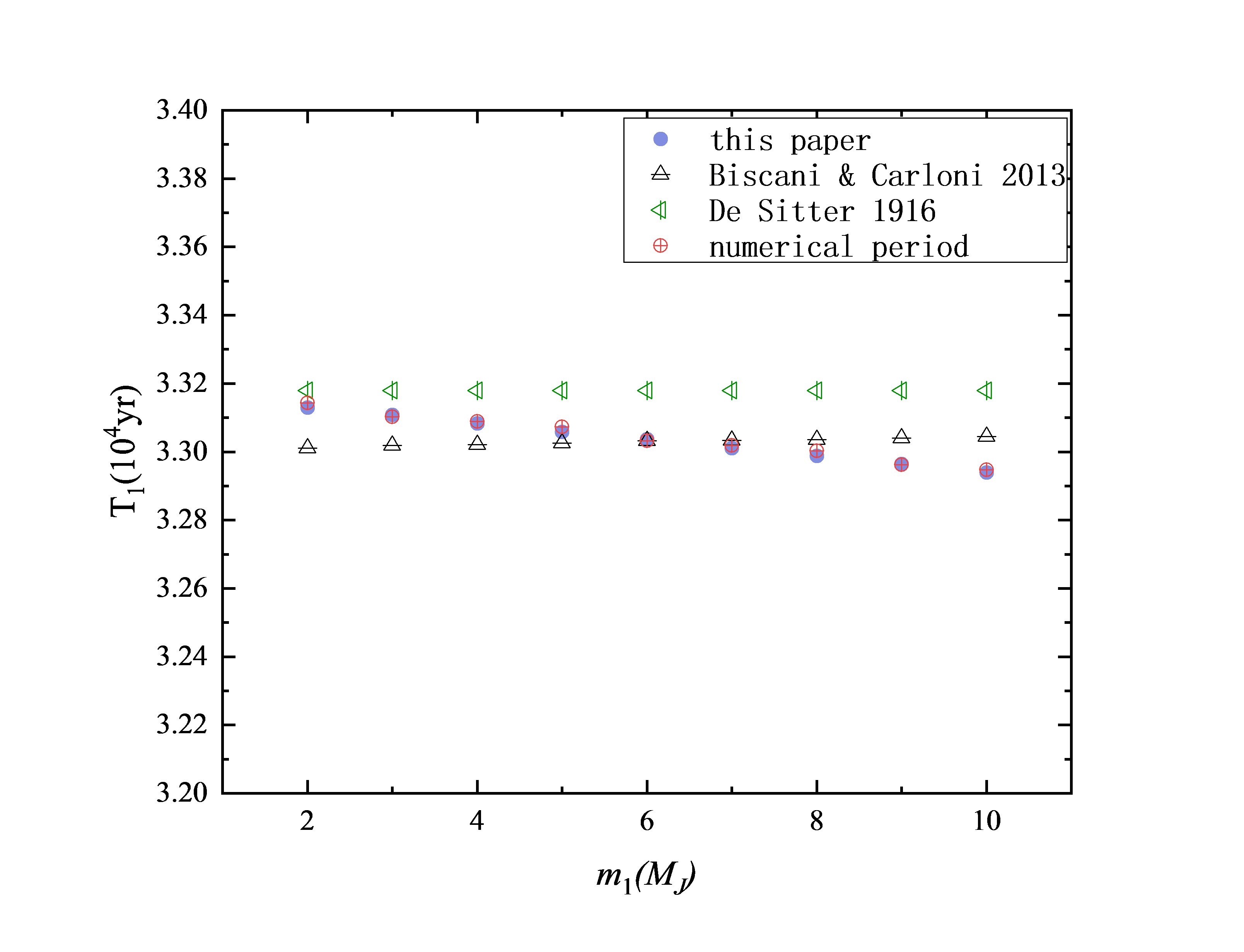}
\includegraphics[width=4in,height=3.4in]{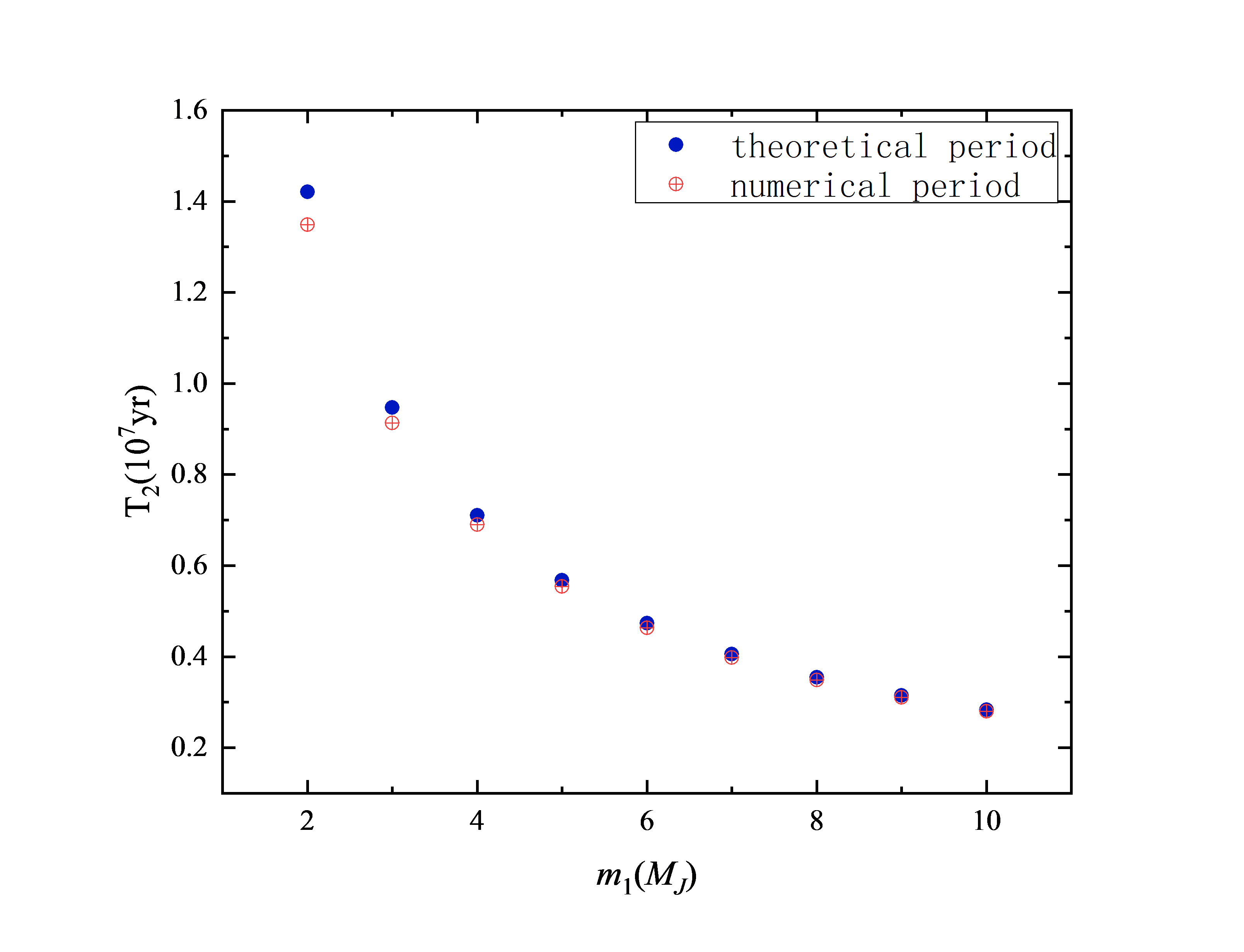}
\caption{The periods of spin precession for both the planet and the host star are simulated numerically, along with the corresponding theoretical results obtained from Equations~\ref{avT1} and \ref{avT2}, with parameters $m_2 = 1.09 M_\odot$, $a = 0.05$ AU, and $e = 0$. Additional theoretical results for the planetary spin precession period were obtained from \citep{Biscani2013} and \citep{DeSitter1916} for comparison. }
\label{periodm}
\end{figure}

\begin{figure}
\epsscale{1.0}
\centering
\includegraphics[width=4in,height=3.4in]{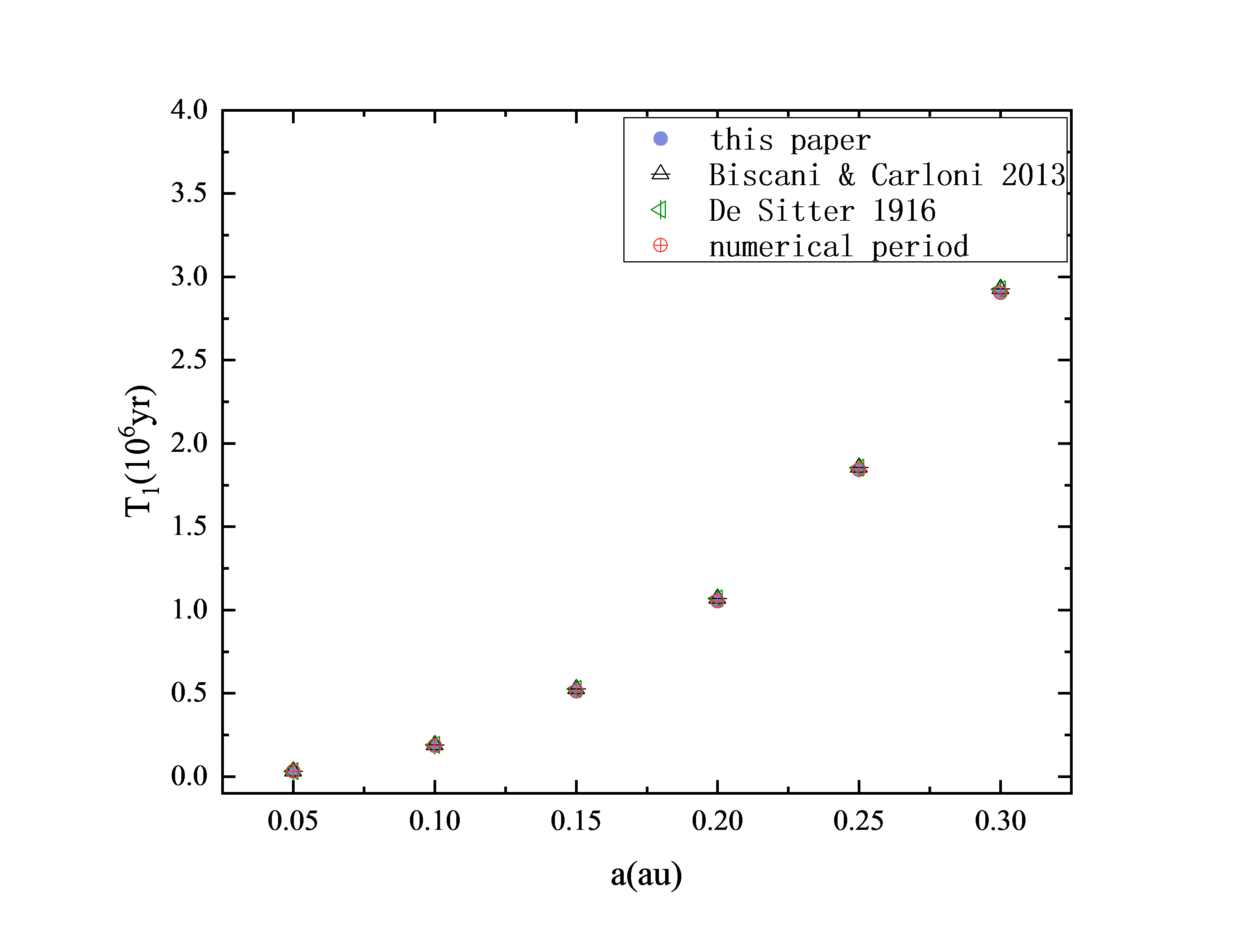}
\includegraphics[width=4in,height=3.4in]{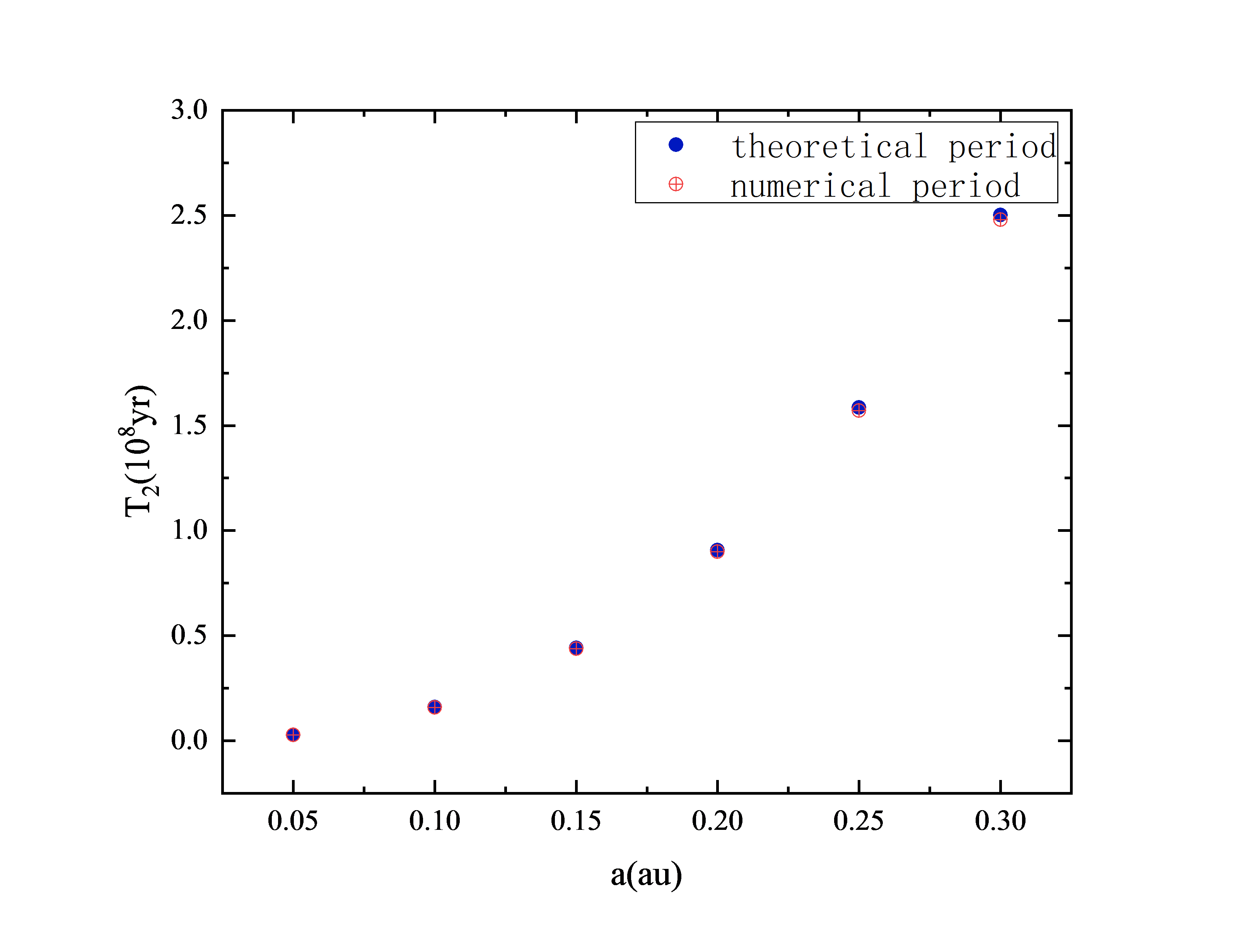}
\caption{The periods of spin precession for both the planet and host star are simulated numerically, along with the corresponding theoretical results obtained from Equations~\ref{avT1} and \ref{avT2}, with parameters $m_1 = 10 M_J$, $m_2 = 1.09 M_\odot$, and $e = 0$. Additional theoretical results for the planetary spin precession period were obtained from \citep{Biscani2013} and \citep{DeSitter1916} for comparison.}
\label{perioda}
\end{figure}

\begin{figure}
\epsscale{1.0}
\centering
\includegraphics[width=4in,height=3.4in]{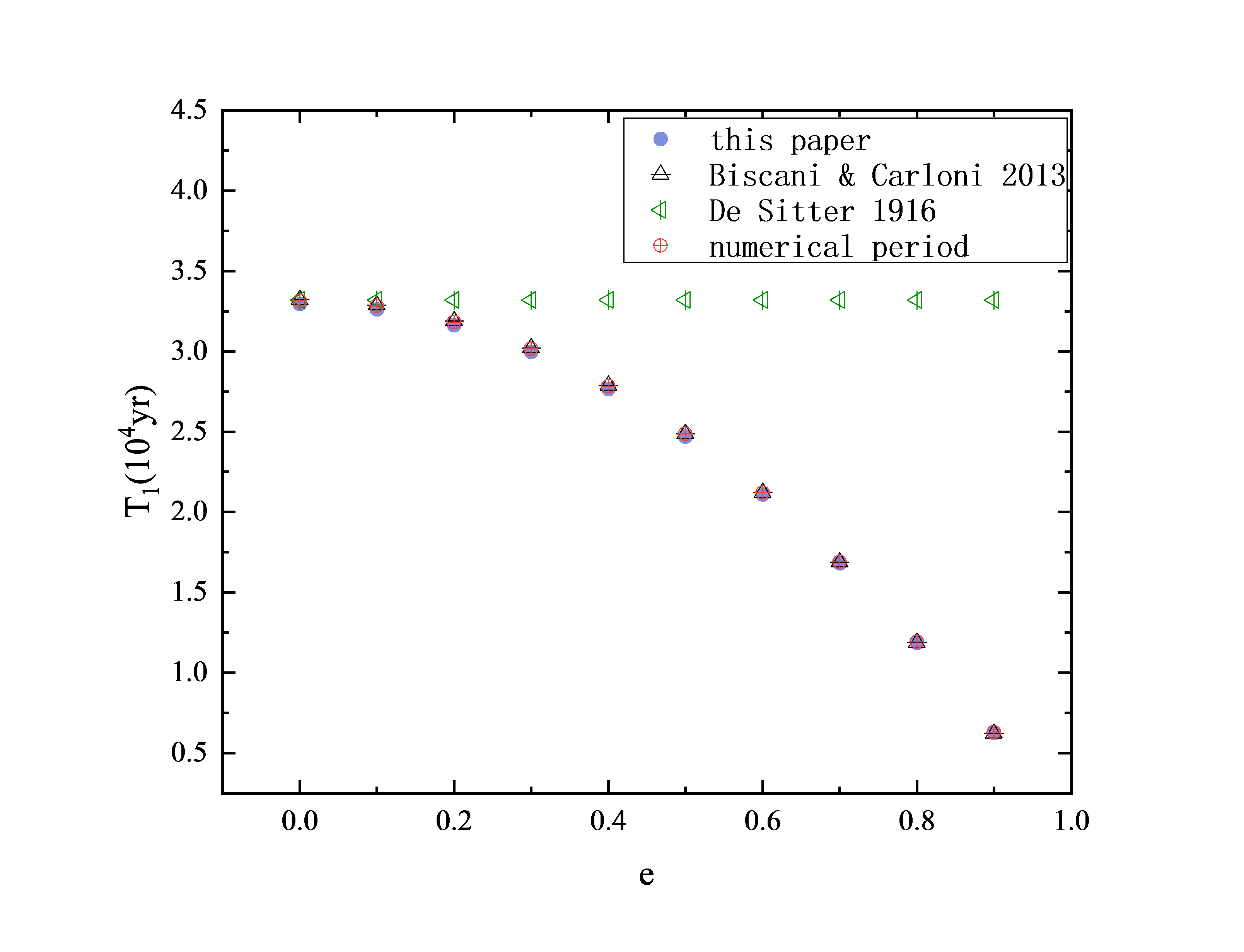}
\includegraphics[width=4in,height=3.4in]{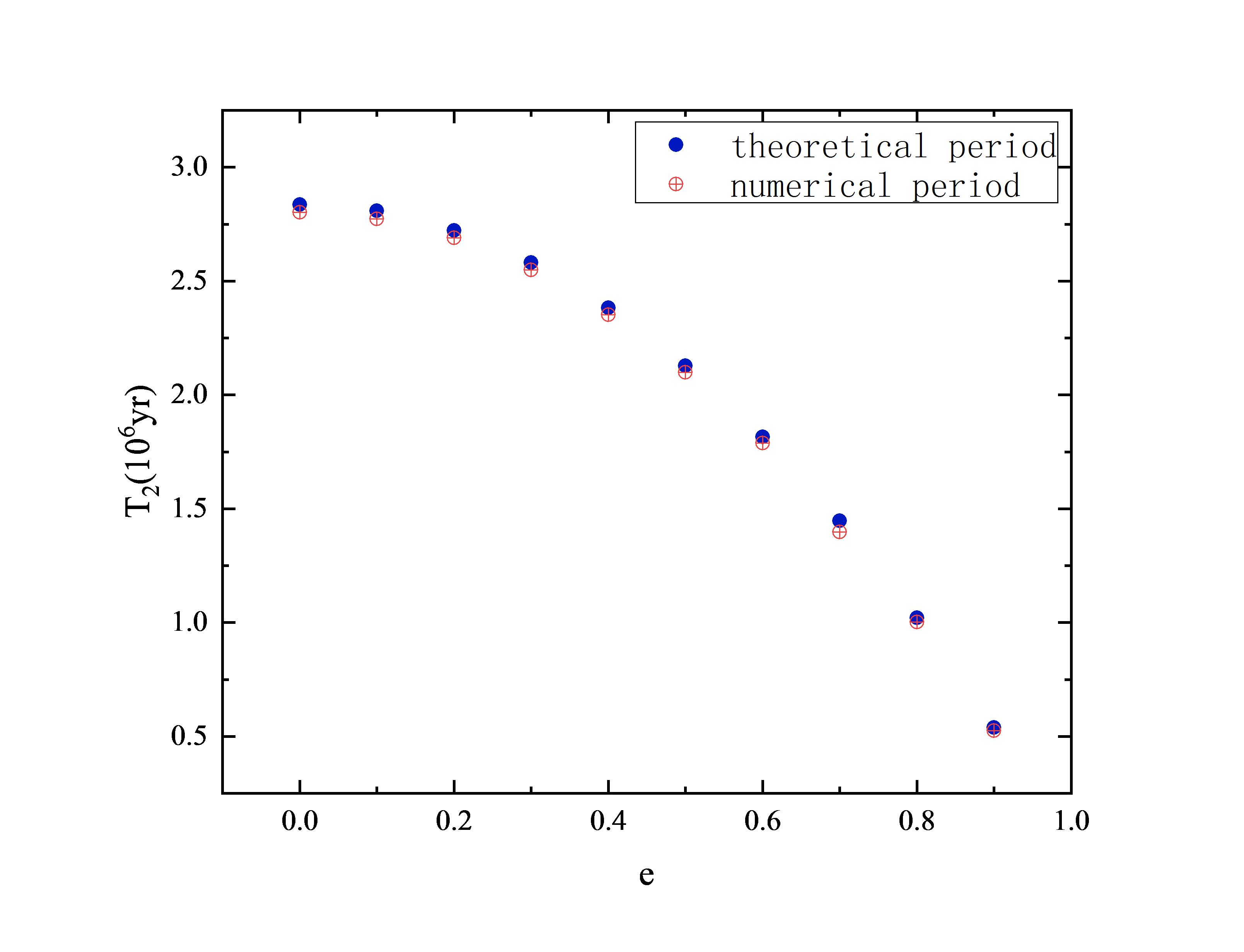}
\caption{The periods of spin precession for both the planet and the host star are computed numerically, alongside the corresponding theoretical results obtained from equations \ref{avT1} and \ref{avT2}, with $m_1=10 M_J$, $m_2=1.09 M_\odot$, and $a=0.05 , \text{au}$. Additional theoretical results for the planetary spin precession period are compared with those from \citep{Biscani2013} and \citep{DeSitter1916}.}
\label{periode}
\end{figure}

From the above equations, we obtain the periods $T_1 = 3.31 \times 10^4$ years and $T_2 = 2.84 \times 10^6$ years, which agree perfectly with the numerical results. We compared the periods of spin precession for both the planet and the host star, as simulated numerically, with the corresponding theoretical results obtained from Equations~\ref{avT1} and \ref{avT2}. The comparisons for varying planetary mass, orbital semi-major axis, and eccentricity are shown in Figures~\ref{periodm}, \ref{perioda}, and \ref{periode}, where we observe good agreement between the theoretical and numerical results.

\section{Influence of 2PN spin-spin coupling term on the periods of spin precession}
    \label{ch.spinspin}
    In this section, we investigate whether the 2PN spin-spin coupling term
    affects the formation of high obliquity of hot Jupiters.
    We incorporate the 2PN spin-spin coupling term to formula (3), so that  $H_1$ is now expressed as:
    \begin{eqnarray}
    H_1=H_{\rm 1PN}+H_{\rm SO}+H_{\rm SS}
    \end{eqnarray}
    Where
    \begin{eqnarray}
    H_{\rm SS}=\frac{G}{r^3} [ 3( \mathbf{J} _1\cdot \mathbf{n})(\mathbf{J} _2\cdot \mathbf{n} )-\mathbf{J}_1\cdot
    \mathbf{J}_2]
    \end{eqnarray}
    
    According to the method of \citep{Wu2010}, the final
    simplified $H_{\rm SS}$ with canonical variables is expressed as:
    \begin{eqnarray}
    H_{\rm SS}&=&\frac{G}{r^3} [\frac{3}{r^2} ((\sqrt{J_1^2-\xi _1^2}\cos \theta _1 )x
    +(\sqrt{J_1^2-\xi _1^2}\sin \theta _1 )\nonumber\\&&y+\xi _1z)
    ((\sqrt{J_2^2-\xi _2^2}\cos \theta _2 )x
    +(\sqrt{J_2^2-\xi _2^2}\sin \theta _2 )\nonumber\\&&y+\xi _2z)-((\sqrt{J_1^2-\xi _1^2}\cos \theta _1 )
    (\sqrt{J_2^2-\xi _2^2}\cos \theta _2 )\nonumber\\&&+(\sqrt{J_1^2-\xi _1^2}\sin \theta _1 )
    (\sqrt{J_2^2-\xi _2^2}\sin \theta _2 )\nonumber\\&&+\xi _1\xi _2)]
    \end{eqnarray}

\begin{figure}
\epsscale{1.0}
\centering
\includegraphics[width=4in,height=3.4in]{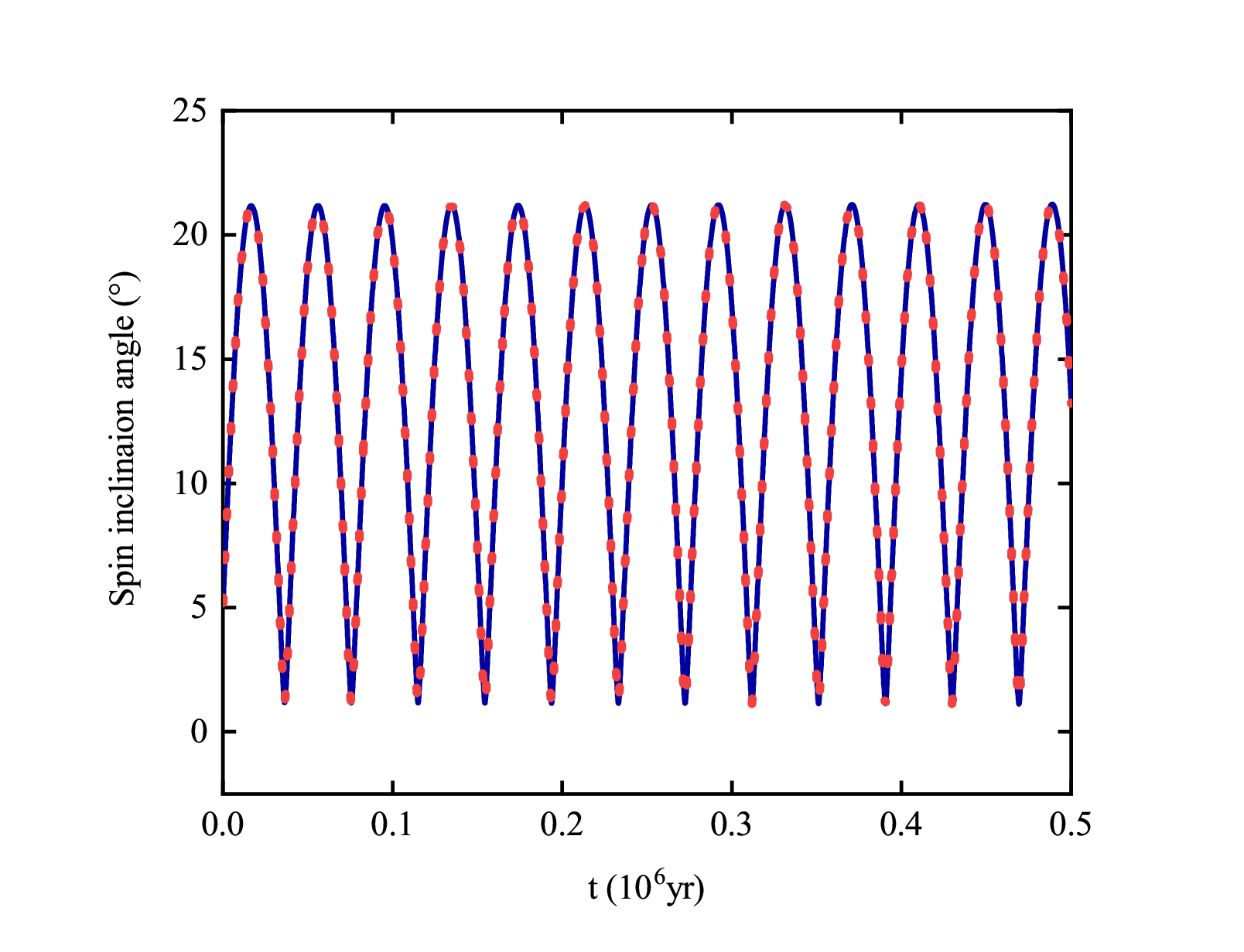}
\caption{The time evolution of the angle between the spin of the hot Jupiter 51 Peg b and the z-axis is shown. The y-axis represents the angle between the spin of 51 Peg b and the z-axis, with time measured in units of $10^6$ years. Blue represents the case without the 2PN spin-spin coupling term, while red corresponds to the case including the 2PN spin-spin coupling term.}
\label{fig5}
\end{figure}

We compare the time evolution of the angle between the planet's spin and the z-axis. The results, shown in Figure \ref{fig5}, reveal that the curves including the 2PN spin-spin coupling term almost overlap with those considering only the 1.5PN spin-orbit coupling term. This indicates that the influence of the spin-spin coupling term on the spin precession periods is negligible, and the 1.5PN spin-orbit coupling term is the dominant factor driving the precession of the spins.

\section{Revelation: Residual planetary disk or exoplanets with obliquity}

\begin{figure}
\epsscale{1.0}
\centering
\includegraphics[width=4in,height=3.4in]{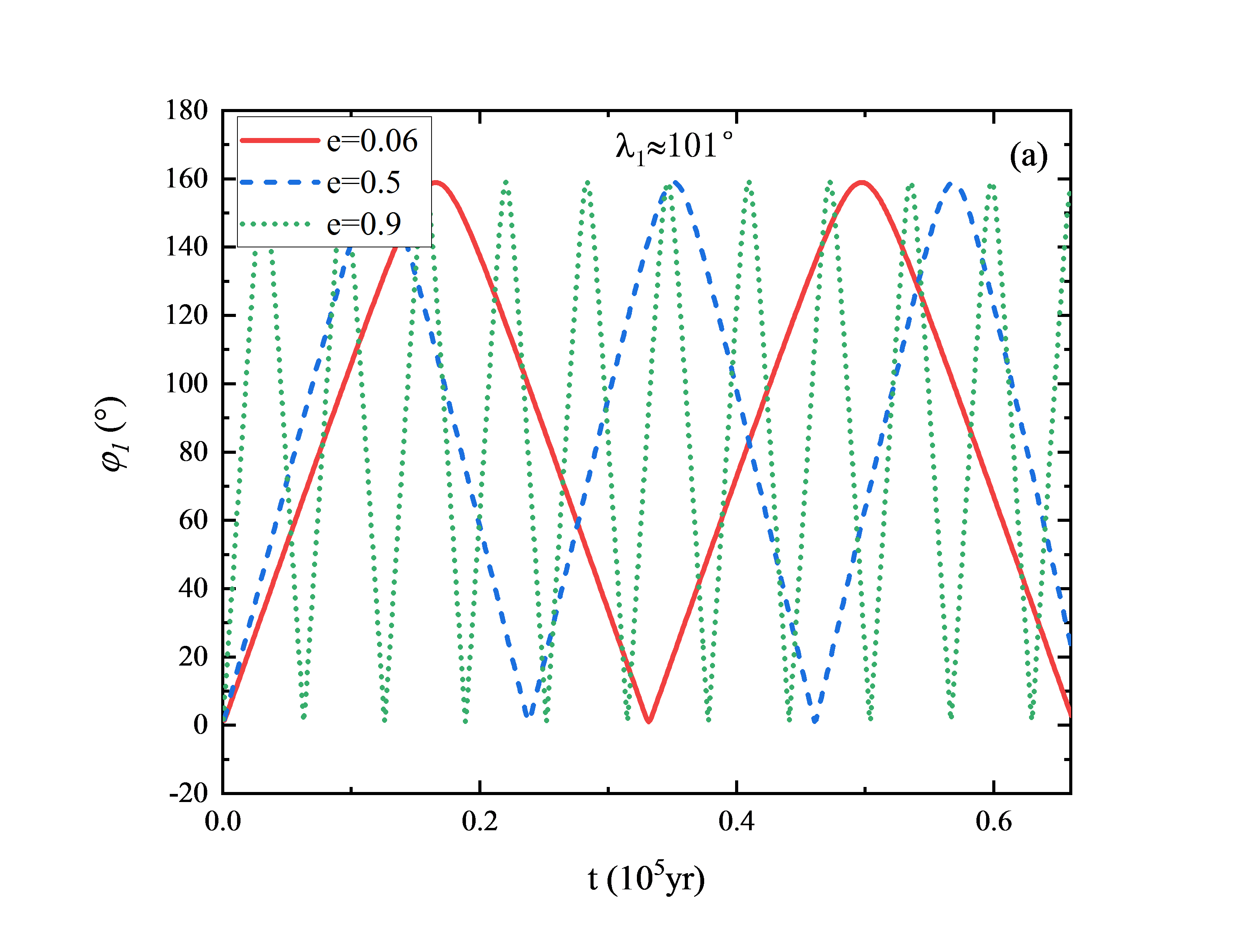}
\includegraphics[width=4in,height=3.4in]{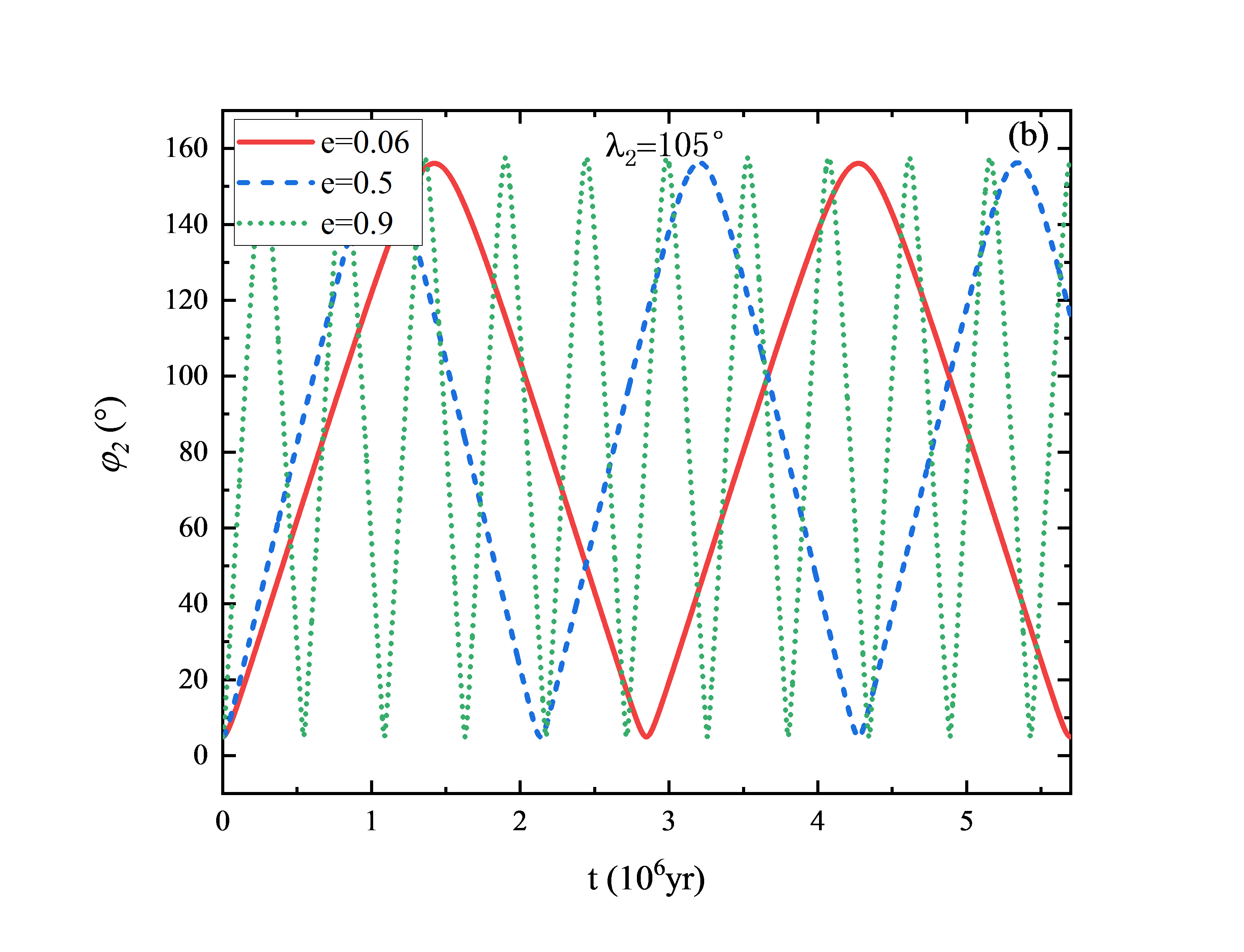}
\caption{ Retrograde planets, $i = 100^\circ$, with different eccentricities $e=0.006,0.5,0.9$ can lead to reversed spins of the planets and the host star. The other orbital parameters are listed in Case-0 of Table \ref{Table2}.  }
\label{orientationretro}
\end{figure}

As is well known, during the dynamical instability stage\citep{Zhou2007, Wang2017, Wang2019}, with the dissipation of the gas disk, planets can be scattered into regions close to the host star. Planets can also migrate inward through disk migration \citep{Goldreich1980, Lin1986, Lin1996, Ward1997}. Due to the post-Newtonian orbital-spin coupling terms, planets migrating inward with small obliquity or inclination can cause the spin orientation of the host star to deviate further from the normal of the planetary disk. Successive similar processes may gradually increase this deviation. Furthermore, a retrograde planetary orbit can reverse the spin orientation of the host star, while more eccentric planetary orbits accelerate the deviation, as shown in Figure \ref{orientationretro}, potentially leaving a residual planetary disk or exoplanets with significant obliquities. The largest excitation timescale is approximately half of the precession period, which can be calculated using Equations \ref{avT1} and \ref{avT2}.

\section{Conclusions}
\label{ch.con}

By utilizing canonical spin variables, the Hamiltonian for a system of two spinning bodies incorporating the post-Newtonian spin-orbit coupling term can be proved integrable. Building on this framework, we analyze the characteristics of spin precession for both the planet and its host star, including their trajectories on a three-dimensional sphere, the amplitude of variation, and their periods, using both numerical simulations and theoretical analysis. The theoretical formula for the period of spin precession is valid without any restrictions and agrees closely with the numerical results. Additionally, the orbital inclination can expand the range of deviation in the spatial orientation of the bodies' spins relative to the plane normal. It is the 1.5PN spin-orbit coupling term, rather than the 2PN spin-spin coupling term, that primarily drives the precession of the spins.





 
    
\end{document}